\documentclass[11pt]{article}
\usepackage[dvips]{color}
\usepackage{epsfig}
\usepackage{amsmath}
\usepackage{graphicx}
\date{}
\begin{document}
\title{{\bf Classical and quantum Ho\v{r}ava-Lifshitz cosmology in a minisuperspace perspective}}
\author{B. Vakili$^{1}$\thanks{b-vakili@iauc.ac.ir}\,\, and\,\,\
V. Kord$^{2}$\thanks{v-kord87@stu.umz.ac.ir}\,\,\,\\\\$^1${\small
{\it Department of Physics, Chalous Branch, Islamic Azad University
(IAU), P.O. Box 46615-397, Chalous, Iran}}\\$^2${\small {\it
Department of Physics, University of Mazandaran, P.O. Box
47416-95447, Babolsar, Iran}}} \maketitle

\begin{abstract}
We study the classical and quantum models of a
Friedmann-Robertson-Walker (FRW) cosmology in the framework of the
gravity theory proposed by Ho\v{r}ava, the so-called
Ho\v{r}ava-Lifshitz theory of gravity. Beginning with the ADM
representation of the action corresponding to this model, we
construct the Lagrangian in terms of the minisuperspace variables
and show that in comparison with the usual Einstein-Hilbert
gravity, there are some correction terms coming from the
Ho\v{r}ava theory. Either in the matter free or in the case when
the considered universe is filled with a perfect fluid, the exact
solutions to the classical field equations are obtained for the
flat, closed and open FRW model and some discussions about their
possible singularities are presented. We then deal with the
quantization of the model in the context of the Wheeler-DeWitt
approach of quantum cosmology to find the cosmological wave
function. We use the resulting wave functions to investigate the
possibility of the avoidance of classical singularities due to
quantum effects.\vspace{5mm}\noindent\vspace{0.8mm}\newline
\end{abstract}
\section{Introduction}
Since $2009$ when a new theory of gravity was first introduced by
Ho\v{r}ava \cite{Hor1}, many efforts have been made in this area
and the corresponding results have been followed by a number of
works, the main motivations of which lie in the results of quantum
gravity and cosmology, see for instance \cite{Gian}. Since this
kind of gravitational theory has its roots in the Lifshitz work on
the second-order phase transition in solid state physics, it is
usually referred to as the Ho\v{r}ava-Lifshitz (HL) theory of
gravity. Like another candidates for quantum gravity such as
string theory, HL gravity is also a completion of general
relativity (GR) at high energy ultraviolet (UV) regime and reduces
to standard GR in the low energy infra-red (IR) limit. However, in
the framework of the HL theory, the well-known phenomenon of
Lorentz symmetry breaking at high energies is described somehow in
a different way. Indeed the basic idea behind HL is that the
Lorentz symmetry will be broken through a Lifshitz-like process,
i.e., through an anisotropic scaling (characterized by a scale
parameter $b$ and the dynamical critical exponent $z$) between
space and time as
\begin{equation}\label{In1}
t\rightarrow b^z t,\hspace{0.5cm}{\bf x}\rightarrow b{\bf x}.
\end{equation}There are theories correspond to the different
values of $z$. While for $z=1$ the standard relativistic scale
invariance with Lorentz symmetry (the IR limit) is recovered, the
UV gravitational theory proposed in \cite{Hor1} requires $z=3$.
Because of the asymmetry of space and time in HL theory the most
common representation of the space-time metric is its ADM form
which in terms of the lapse function $N(t,{\bf x})$, shift vector
$N^a(t,{\bf x})$ and spacial metric $h_{ab}(t,{\bf x})$ takes the
general form
\begin{equation}\label{In2}
g_{\mu \nu}(t,{\bf x})=\left(%
\begin{array}{cc}
-N^2(t,{\bf x})+N_a(t,{\bf x})N^a(t,{\bf x}) & N_b(t,{\bf x}) \\
N_a(t,{\bf x}) & h_{ab}(t,{\bf x}) \\
\end{array}%
\right).
\end{equation}Depending on whether the lapse function is a function
only of $t$ or of $(t,{\bf x})$, the theory is called projectable or
non-projectable. As pointed out in \cite{Hor1}, since in
cosmological metrics the lapse function labels the time parameter,
it is quite reasonable to choose it as a projectable function.
However, more general cases in which the lapse function is taken as
a non-projectable function are studied in \cite{Blas}. Indeed, in
these references the authors have examined some consistency
conditions present in the projectable HL theory and its
non-projectable extension. They found that while the projectable
theory suffers from the presence of the ghost mode and hence cannot
be consistent, the non-projectable models are free from such ghost
instabilities. As a remark we would like to to emphasize that this
problem cannot be seen in the minisuperspace approximation of
cosmological models in which we restrict the metric and the matter
fields to be homogeneous. Therefore, the study of the minisuperspace
cosmology in the framework of projectable HL gravity is quite
reasonable. However, one should bear in mind that even at the
classical level the above mentioned instability could be visible
when the the perturbations analysis about homogeneous background is
performed. There is yet another issue related to the HL theory, the
so called {\it detailed balance condition}. The gravitational action
of the model consists of the kinetic part ${\cal S}_K$ and the
potential part ${\cal S}_V$ as ${\cal S}={\cal S}_K+{\cal S}_V$. The
kinetic part comes from the Einstein-Hilbert action usually written
in terms of the ADM variables (see the relation (\ref{B}) bellow).
For the potential part the following general form is proposed in
\cite{Hor1}
\begin{equation}\label{In3}
{\cal S}_V=\int d^4x\sqrt{-g}V[h_{ab}],
\end{equation}where $V$ is a scalar function which depends only on
the spacial metric $h_{ab}$ and its spacial derivatives. Taking a
three dimensional HL model with $z=3$, this function may be
constructed by a superposition of the quadratic (in curvature)
terms such as
\begin{equation}\label{In4}
\nabla_a R_{bc}\nabla^a R^{bc},\hspace{0.5cm}\nabla_a
R_{bc}\nabla^b
R^{ca},\hspace{0.5cm}R\nabla^2R,\hspace{0.5cm}R^{ab}\nabla^2
R_{ab},
\end{equation}and cubic terms such as
\begin{equation}\label{In4}
R^3,\hspace{0.5cm}R^a_b R^b_c R^c_a,\hspace{0.5cm}RR_{ab}R^{ab}.
\end{equation}Among the very different possible combinations,
Ho\v{r}ava considered a special form, known as "detailed balance
condition" satisfying model as
\begin{equation}\label{In5}
{\cal S}_V \sim \int d^4x \sqrt{-g} E^{ab}{\cal G}_{abcd}E^{cd},
\end{equation}where ${\cal G}_{abcd}$ is the DeWitt metric and
$E^{ab}$ is a tensor constructed by variation of some function
$W[h_{ab}]$ with respect to the spacial metric. By choosing a
suitable ansatz for the function $W$, he then showed that in a
$z=3$ theory with detailed balance condition the potential is a
combination of the terms (see \cite{Hor1} for details)
\begin{equation}\label{In6}
\nabla_a R_{bc} \nabla^a R^{bc},\hspace{0.5cm}\nabla_a
R_{bc}\nabla^b R^{ac},\hspace{0.5cm}\nabla_a R \nabla^a R.
\end{equation}The issue of the detailed balance
condition makes the theory to have simpler renormalization
properties. Although the detailed balanced system exhibit simpler
quantum behavior, it is shown in \cite{Sot} that if one relaxes
it, the resulting Lagrangian with extra allowed terms is
well-behavior enough to recover the model with detailed balance.

In this paper we shall consider a FRW cosmological model in the
framework of a projectable HL gravity without detailed balance
condition. Our approach to deal with such a problem is through its
representation with minisuperspace variables. Minisuperspace
formulation of HL cosmology is studied in some works, see for
instance \cite{Ber}-\cite{Chris}, to obtain its possible classical
and quantum solutions. Here, we first consider the vacuum case in
which the cosmological model is free from the presence of any kind
of matter fields and see that the corresponding classical
solutions exhibit some types of singularities. We then will add a
perfect fluid as the matter into the model and obtain some exact
solutions in the cases of flat, closed and open FRW cosmologies.
Since our aim in the quantum part of the model is to investigate
the time evolution of the wave function, we prefer to use the
perfect fluid in its Schutz formalism in which the Hamiltonian of
the fluid consists of a linear momentum, the variable canonically
conjugate to which may play the role of a time parameter. In both
vacuum and perfect fluid cases, we construct the corresponding
quantum cosmology based on the canonical approach of
Wheeler-DeWitt theory to see how things may change their behavior
if the quantum mechanical considerations come into the model.
\section{The model}
In this section we start by the FRW cosmology within the framework
of HL gravity. In a quasi-spherical polar coordinate the metric of
space time is assumed to be
\begin{equation}\label{A}
ds^2=-N^2(t)dt^2+a^2(t)\left[\frac{dr^2}{1-kr^2}+r^2\left(d\vartheta^2+\sin^2\vartheta
d\varphi\right)\right],\end{equation}where $N(t)$ is the lapse
function, $a(t)$ the scale factor and $k=1$, $0$ and $-1$
corresponds to the closed, flat and open universe respectively. It
is clear that in terms of the ADM variables the above metric can
be written as
\begin{equation}\label{A1}
ds^2=-N^2(t)dt^2+h_{ab}dx^adx^b,\end{equation}in which
\[h_{ab}=a^2(t)\mbox{diag}\left(\frac{1}{1-kr^2},r^2,r^2\sin^2\vartheta\right),\]is
the intrinsic metric induced on the spatial $3$-dimensional
hypersurfaces. The action of the model consists of the
gravitational part ${\cal S}_g$ and the matter action ${\cal S}_m$
as
\begin{equation}\label{A2}
{\cal S}={\cal S}_g+{\cal S}_m.\end{equation}The matter part of
the action is independent of the HL corrections to the gravity
part. In the context of the ADM formalism, following \cite{Sot}
and with the same notation as is used in \cite{Ber,Paulo}, the
projectable HL gravity without detailed balance has the
(gravitational) action (in what follows we work in units where
$c=\hbar=1$)
\begin{eqnarray}\label{B}
{\cal S}_{g}=\frac{M_{Pl}^2}{2}\int_{{\cal M}} d^3xdt N
\sqrt{h}\left[K_{ab}K^{ab}-\lambda K^2-g_0
M_{Pl}^2-g_1R-M_{Pl}^{-2}\left(g_2R^2+g_3
R_{ab}R^{ab}\right)\nonumber \right.\\ \left.
-M_{Pl}^{-4}\left(g_4 R^3 +g_5 R R^a_b R^b_a+g_6 R^a_b R^b_c
R^c_a+g_7 R \nabla ^2 R +g_8 \nabla_a R_{bc}\nabla^a
R^{bc}\right)\right]+M_{Pl}^2\int_{\partial {\cal M}} d^3x
\sqrt{h}K,
\end{eqnarray}where $M_{Pl}=(8\pi G)^{-1/2}$ is the Planck mass, $K_{ab}$ are the components of the extrinsic curvature tensor
which describes how much the spatial space $h_{ab}$ (which is the
boundary $\partial{\cal M}$ of the four-dimensional manifold
${\cal M}$) is curved in the way it sits in the space time
manifold. Also, $h$ and $R$ are the determinant and Ricci scalar
of the spatial geometry $h_{ab}$ respectively, and $K$ represents
the trace of $K_{ab}$. The constants $\lambda$ and $g_i$
($i=0,1,...,8$) denote the HL corrections to the usual GR. In
comparison with the ADM representation of GR action, that is
\[{\cal S}_g=\frac{M_{Pl}^2}{2}\left[\int_{{\cal M}} d^3x dt N
\sqrt{h}\left(K_{ab}K^{ab}-K^2+R-2\Lambda\right)+2\int_{\partial
{\cal M}} d^3x \sqrt{h}K\right],\]we conclude that the consensus
GR is recovered if one sets the cosmological constant as
$\Lambda=g_0M_{Pl}^2/2$, $g_1=-1$ and $\lambda=1$. However, as is
mentioned in \cite{Ber}, we take $\lambda$ as a running constant
which represents the IR limit of the gravitational theory. With
these identifications, the action (\ref{B}) may be rewritten as
\begin{eqnarray}\label{C}
{\cal S}_{g}=\frac{M_{Pl}^2}{2}\int_{{\cal M}} d^3xdt N
\sqrt{h}\left[K_{ab}K^{ab}-\lambda
K^2+R-2\Lambda-M_{Pl}^{-2}\left(g_2R^2+g_3
R_{ab}R^{ab}\right)\nonumber \right.\\ \left.
-M_{Pl}^{-4}\left(g_4 R^3 +g_5 R R^a_b R^b_a+g_6 R^a_b R^b_c
R^c_a+g_7 R \nabla ^2 R +g_8 \nabla_a R_{bc}\nabla^a
R^{bc}\right)\right]+M_{Pl}^2\int_{\partial {\cal M}} d^3x
\sqrt{h}K.
\end{eqnarray}
From its standard definition, the extrinsic curvature is given by
\begin{equation}\label{D}
K_{ab}=\frac{1}{2N}\left(N_{a|b}+N_{b|a}-\frac{\partial
h_{ab}}{\partial t}\right),\end{equation}where $N_a$ is the shift
vector and $N_{a|b}$ represents the covariant derivative with
respect to $h_{ab}$. Since the shift vector is absent in the FRW
models, a simple calculation based on the above definition results
in $K_{ab}K^{ab}=\frac{3\dot{a}^2}{N^2 a^2}$ and
$K=-\frac{3\dot{a}}{Na}$, where a dot represents differentiation
with respect to $t$. Also, the Ricci tensor and the Ricci scalar
correspond to the $3$-geometry $h_{ab}$ can be obtained as
$R_{ab}=\frac{2kh_{ab}}{a^2}$ and $R=\frac{6k}{a^2}$. The
gravitational part for FRW model may now be written by
substituting the above results into action (\ref{C}), giving
\begin{eqnarray}\label{E}
{\cal S}_g=\frac{3V_0 M_{Pl}^2(3\lambda-1)}{2}\int dt N
\left\{-\frac{a\dot{a}^2}{N^2}+\frac{6ka}{3(3\lambda-1)}-\frac{2\Lambda
a^3}{3(3\lambda-1)}-M_{pl}^{-2}\left[\frac{12k^2(3g_2+g_3)}{3a(3\lambda-1)}\right]\nonumber
\right.\\ \left.
-M_{Pl}^{-4}\left[\frac{24k(9g_4+3g_5+g_6)}{3a^3(3\lambda-1)}\right]\right\},
\end{eqnarray}where $V_0=\int d^3x\sqrt{h}$ is the integral over
spatial dimensions.  If we set $\frac{3V_0
M_{Pl}^2(3\lambda-1)}{2}=1$, then we are led to the point-like
Lagrangian
\begin{equation}\label{F}
{\cal
L}_g=N\left(-\frac{a\dot{a}^2}{N^2}+g_cka-g_{\Lambda}a^3-\frac{g_rk^2}{a}-\frac{g_sk}{a^3}\right),\end{equation}where
its coefficients are defined as \cite{Kei}
\begin{equation}\label{G}
g_c=\frac{2}{3\lambda-1},\hspace{0.5cm}g_{\Lambda}=\frac{2\Lambda}{3(3\lambda-1)},\hspace{0.5cm}g_r=6V_0(3g_2+g_3),\hspace{0.5cm}
g_s=18V_0^2(3\lambda-1)(9g_4+3g_5+g_6).\end{equation}Now, the
gravitational part of the Hamiltonian for this model can be
obtained from its standard definition $H_g=\dot{a}P_a-{\cal L}_g$.
Noting that
\begin{equation}\label{H}
P_a=\frac{\partial {\cal L}_g}{\partial
\dot{a}}=-2\frac{a\dot{a}}{N},\end{equation}one gets
\begin{equation}\label{I}
H_g=N{\cal
H}_g=N\left[-\frac{P_a^2}{4a}-g_cka+g_{\Lambda}a^3+\frac{g_rk^2}{a}+\frac{g_sk}{a^3}\right].\end{equation}We
see that the lapse function enters in the Hamiltonian as a
Lagrange multiplier, as expected. Thus, when we vary the
Hamiltonian with respect to $N$, we get ${\cal H}_g=0$, which is
called the Hamiltonian constraint. On the classical level this
constraint is equivalent to the Friedmann equation, while on the
quantum level, the operator version of this constraint annihilates
the wave function of the corresponding universe, leading to the
so-called Wheeler-DeWitt equation. Now, let us deal with the
matter field with which the action of the model is augmented. As
we have mentioned, the matter part of the action is independent of
modifications due to the HL terms. Therefore, the matter may come
into play in a common way and the total Hamiltonian can be made by
adding the matter Hamiltonian to the gravitational part (\ref{I}).
To do this, we consider a perfect fluid whose pressure $p$ is
linked to its energy density $\rho$ by the equation of state
\begin{equation}\label{J}
p=\omega \rho,\end{equation}where $-1\leq \omega \leq1$ is the
equation of state parameter. According to Schutz's representation
for the perfect fluid \cite{Schutz}, its Hamiltonian can be viewed
as (see \cite{Vak} for details)
\begin{equation}\label{L}
H_m=N\frac{P_T}{a^{3\omega}},\end{equation}where $T$ is a
dynamical variable related to the thermodynamical parameters of
the perfect fluid and $P_T$ is its conjugate momentum. Finally, we
are in a position in which can write the total Hamiltonian $H=H_g+
H_m$ as
\begin{equation}\label{M}
H=N{\cal
H}_g=N\left[-\frac{P_a^2}{4a}-g_cka+g_{\Lambda}a^3+\frac{g_rk^2}{a}+\frac{g_sk}{a^3}+\frac{P_T}{a^{3\omega}}\right].\end{equation}The
setup for constructing the phase space and writing the Lagrangian
and Hamiltonian of the model is now complete. In the following
sections, we shall deal with classical and quantum cosmologies
which can be extracted from a theory with the above mentioned
Hamiltonian.
\section{Cosmological dynamics without matter}
\subsection{Classical model}
\begin{figure}
\begin{tabular}{ccc} \epsfig{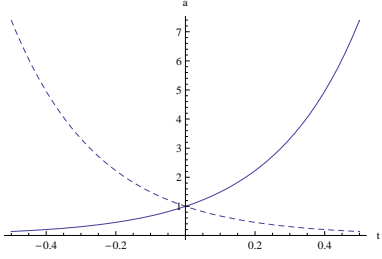}
\hspace{1cm}
\end{tabular}
\caption{\footnotesize Qualitative behavior of the scale factor
for the flat universe. Solid and dashed lines denote the relation
(\ref{O}) for positive and negative signs
respectively.}\label{fig1}
\end{figure}

When the model is free of the contribution of any kind of matter
fields, the dynamics is described by Hamiltonian (\ref{I}).
Therefore, the classical equations of motion is governed by the
Hamiltonian equations. Equivalently, we can use the constraint
equation ${\cal H}=0$ which is nothing but the variation the
action with respect to $N$. In this sense, from (\ref{H}) and
(\ref{I}) we obtain
\begin{equation}\label{N}
a\dot{a}^2+g_c k a-g_{\Lambda}a^3-\frac{g_r k^2}{a}-\frac{g_s
k}{a^3}=0.\end{equation} For the flat case $k=0$, there is no
difference between the HL cosmology and the usual FRW model except
that the cosmological constant shifts as $\Lambda \rightarrow
3g_{\Lambda}=\frac{2\Lambda}{3\lambda-1}$. In this case the above
equation admits the following solutions
\begin{equation}\label{O}
a(t)=a_0e^{\pm \sqrt{g_{\Lambda}} t},\end{equation}where $a_0$ is
an integration constant and the positive (negative) sign in the
power of the exponential function corresponds to the expansion
(contraction) universe. Thus, in this case we have two distinct
branches of solutions, one of which begins with zero size at
$t=-\infty$ and expands forever according to an exponential de
Sitter law, while another has an opposite behavior, i.e.,
decreases its size from the large values at $t=-\infty$ and tends
exponentially to zero as $t\rightarrow \infty$. The situation of
the dynamical behavior of the scale factor in this case is shown
in figure \ref{fig1}.

\begin{figure}
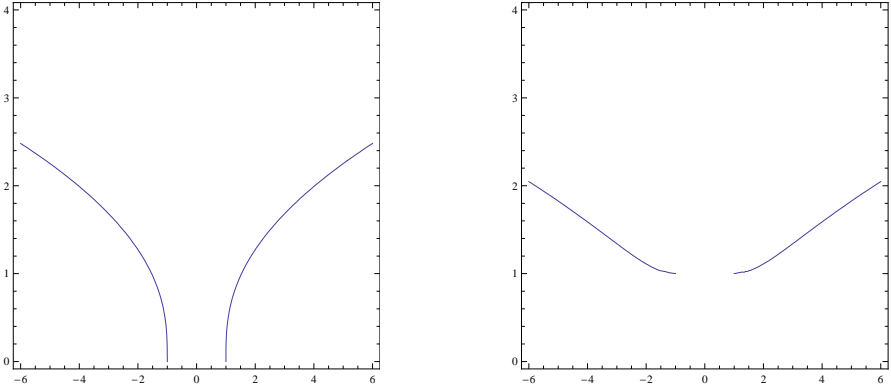

\begin{tabular}{c}\epsfig{figure=Fig2.eps,width=5cm}
\hspace{1.5cm} \epsfig{figure=Fig3.eps,width=5cm}
\end{tabular}
\caption{\footnotesize  The figures show the evolutionary behavior
of the non-flat universe in the early times based on (\ref{Q}).
The left and right figures correspond to the $k=1$ and $k=-1$
respectively.}\label{fig2}
\end{figure}

For $k\neq 0$, equation (\ref{N}) does not seem to have analytical
solutions. So we consider the behavior of its solutions only for
some limiting cases. When the scale factor is very small, we keep
the two last terms in (\ref{N}) and rewrite it as
\begin{equation}\label{P}
a\dot{a}^2-\frac{g_r k^2}{a}-\frac{g_s
k}{a^3}=0,\end{equation}from which we obtain the following
implicit relation between $t$ and $a$
\begin{equation}\label{Q}
\frac{a\sqrt{g_r a^2+kg_s}}{2g_r}-\frac{kg_s\ln\left(g_r
a+\sqrt{g_r^2a^2+kg_s}\right)}{2g_r^{3/2}}=\pm
t-t_0,\end{equation}where $t_0$ is an integration constant. In
figure \ref{fig2}, we have plotted the scale factor versus time
based on the above relation.

\begin{figure}
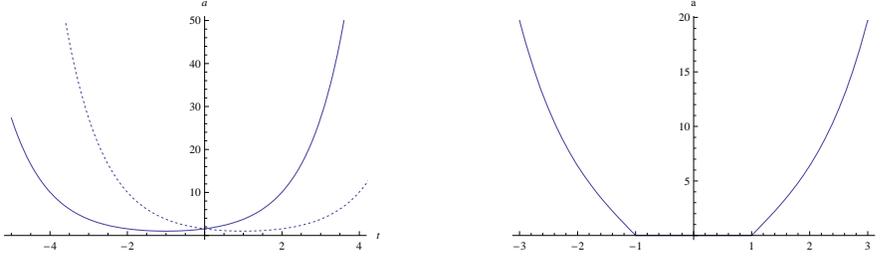

\begin{tabular}{c}\epsfig{figure=Fig4.eps,width=5cm}
\hspace{1.5cm} \epsfig{figure=Fig5.eps,width=5cm}
\end{tabular}
\caption{\footnotesize Left: Late time behavior of the non-flat
universe for $k=1$. Right: The same figure for $k=-1$.
}\label{fig3}
\end{figure}

On the other hand, for the late time of cosmic evolution which the
scale factor is expected to be large the terms with coefficients
$g_c$ and $g_{\Lambda}$ in (\ref{N}) become more important. In
such a case we write this equation as
\begin{equation}\label{R}
\dot{a}^2+g_c k-g_{\Lambda}a^2=0.\end{equation}For $k=+1$, this
equation has two sets of solutions as
\begin{equation}\label{S}
a_I(t)=\frac{1}{2g_{\Lambda}}\left[e^{\sqrt{g_{\Lambda}}(t+t_0)}+g_cg_{\Lambda}e^{-\sqrt{g_{\Lambda}}(t+t_0)}\right],
\end{equation}and
\begin{equation}\label{T}
a_{II}(t)=\frac{1}{2g_{\Lambda}}\left[e^{-\sqrt{g_{\Lambda}}(t-t_0)}+g_cg_{\Lambda}e^{\sqrt{g_{\Lambda}}(t-t_0)}\right],
\end{equation}where $t_0$ is an integration constant. Each of
these solutions consists of two branches. In one branch the
universe contracts and when reaches a minimum size undergoes to an
expansion period. Therefore, in the case of of a closed universe
we have bouncing cosmologies in which the bounce occurs at
$t=-[t_0+\frac{1}{2\sqrt{g_{\Lambda}}}\ln
(\frac{1}{g_cg_{\Lambda}})]$ for $a_I(t)$ and at
$t=[t_0-\frac{1}{2\sqrt{g_{\Lambda}}}\ln (g_cg_{\Lambda})]$ for
$a_{II}(t)$. On the other hand, for $k=-1$ the solutions to the
equation (\ref{R}) take the form
\begin{eqnarray}\label{U}
a(t)=\left\{
\begin{array}{ll}
a^I(t)=\frac{1}{2g_{\Lambda}}\left[e^{-\sqrt{g_{\Lambda}}(t-t_0)}-g_c
g_{\Lambda}e^{\sqrt{g_{\Lambda}}(t-t_0)}\right],\hspace{0.5cm}t<t_0-
\frac{1}{\sqrt{2g_{\Lambda}}}\ln(g_c g_{\Lambda}),\\\\
a^{II}(t)=\frac{1}{2g_{\Lambda}}\left[e^{\sqrt{g_{\Lambda}}(t+t_0)}-g_c
g_{\Lambda}e^{-\sqrt{g_{\Lambda}}(t+t_0)}\right],\hspace{0.5cm}t>-t_0+
\frac{1}{\sqrt{2g_{\Lambda}}}\ln(g_c g_{\Lambda}).
\end{array}\right.
\end{eqnarray}We see that in the time interval $t<t_0-
\frac{1}{\sqrt{2g_{\Lambda}}}\ln(g_c g_{\Lambda})$ the scale
factor decrease and tends its evolution at $t_0-
\frac{1}{\sqrt{2g_{\Lambda}}}\ln(g_c g_{\Lambda})$ with zero size.
Also, for $t>-t_0+ \frac{1}{\sqrt{2g_{\Lambda}}}\ln(g_c
g_{\Lambda})$, the universe has an expanding behavior begins its
evolution with a zero size singularity at $-t_0+
\frac{1}{\sqrt{2g_{\Lambda}}}\ln(g_c g_{\Lambda})$. These two
branches of solutions are separated from each other by a
classically forbidden region corresponding to the time interval
$t_0- \frac{1}{\sqrt{2g_{\Lambda}}}\ln(g_c g_{\Lambda})<t<-t_0+
\frac{1}{\sqrt{2g_{\Lambda}}}\ln(g_c g_{\Lambda})$ for which no
open classical solutions exist. The above results are summarized
in figure \ref{fig3}.

\subsection{Quantum model}
\begin{figure}
\begin{tabular}{ccc} \epsfig{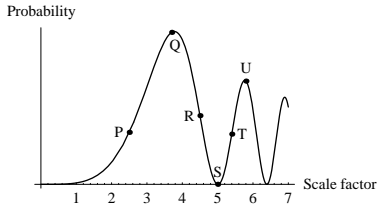}
\hspace{1cm}
\end{tabular}
\caption{\footnotesize Qualitative behavior of the probability
($=|\Psi(a)|^2$) versus scale factor for the flat empty universe,
see relation (\ref{X}).}\label{fig4}
\end{figure}

Now, we shall study the quantum behavior of the model described by
the Hamiltonian (\ref{I}). One way to do such a study is to
investigate the well-known Wheeler-DeWitt equation for the
corresponding universe, that is
\begin{equation}\label{V}
\left[\frac{1}{a}\frac{d^2}{da^2}-\frac{p}{a^2}\frac{d}{da}+4\left(-g_cka+g_{\Lambda}a^3+\frac{g_rk^2}{a}+\frac{g_sk}{a^3}\right)\right]\Psi(a)=0,
\end{equation}where $\Psi(a)$ is the wave function corresponds to
the quantum universe and the parameter $p$ represents the ambiguity
in the ordering of factors $a$ and $P_a$ in the first term of
(\ref{I}). In the case of a flat universe $k=0$, to see the
correspondence of the classical and quantum solutions, we note that
for large values of scale factor, the behavior of the system can be
obtained in the WKB (semiclassical) approximation. Then substituting
$\Psi(a)=\Omega(a)e^{iS(a)}$ in equation (\ref{V}) leads to the
modified Hamilton-Jacobi equation
\begin{equation}\label{Y}
-\frac{1}{4a}\left(\frac{ds}{da}\right)^2+g_{\Lambda}a^3+{\cal
Q}(a)={\cal H}\left(a,P_a=\frac{dS}{da}\right)+{\cal
Q}(a)=0,\end{equation}in which the quantum potential is defined as
${\cal Q}(a)=\frac{1}{4a\Omega}\frac{d^2\Omega}{da^2}-\frac{1}{a^2
\Omega}\frac{d\Omega}{da}$. It is well known that the quantum
effects are important for small values of the scale factor and in
the limit of the large scale factor can be neglected. Therefore,
in the semiclassical approximation region we can omit the ${\cal
Q}$ term in equation (\ref{Y}) and obtain the phase function
$S(a)$ as $S(a)=\pm \frac{2}{3}\sqrt{g_{\Lambda}}a^3$. In the WKB
method, the correlation between classical and quantum solutions is
given by the relation $P_a=\frac{\partial S}{\partial a}$. Thus,
using the definition of $P_a$ in (\ref{H}), the equation for the
classical trajectories becomes $a(t)=a_0e^{\pm
\sqrt{g_{\Lambda}}t}$, which shows that the classical cosmology of
equation (\ref{O}) is exactly recovered. The meaning of this
result is that for large values of the scale factor, the effective
action corresponding to the expanding and contracting universes is
very large and the universe can be described classically. On the
other hand, for small values of the scale factor we cannot neglect
the quantum effects and the classical description breaks down.
Since the WKB approximation is no longer valid in this regime, one
should go beyond the semiclassical approximation. For the flat FRW
metric we have $k=0$ and the two linearly independent solutions to
equation (\ref{V}) can be expressed in terms of the Bessel
functions $J_{\nu}(x)$ leading to the following general solution
\begin{equation}\label{X}
\Psi(a)=a\left[c_1J_{1/3}\left(\frac{2}{3}\sqrt{g_{\Lambda}}a^3\right)+c_2J_{-1/3}\left(\frac{2}{3}\sqrt{g_{\Lambda}}a^3\right)\right],
\end{equation}where $c_{1,2}$ are integration constants. To find the above solutions we have noted that
the factor-ordering parameter $p$ does not affect the semiclassical
probabilities \cite{Vil}, and so we have chosen $p=1$ to make the
differential operator appearing in the Wheeler-DeWitt equation the
Laplacian operator of the minisupermetric (see \cite{Ch}). With an
eye to the classical solutions (\ref{O}), it is clear that the model
is free of classical Big-Bang singularity in which $a(t_0)=0$ for
some $t_0$. Therefore, we expect that the quantum model predicts a
zero probability for the creation of the universe with zero size.
This fact lead us to impose the boundary condition $\Psi(a=0)=0$ on
the wave function, which results in $c_2=0$. Note that equation
(\ref{V}) is a Schr\"{o}dinger-like equation for a fictitious
particle with zero energy moving in the field of the superpotential
$U(a)=-4g_{\Lambda}a^4$. Usually, in the presence of such a
potential, the mini-superspace can be divided into two regions,
$U>0$ and $U<0$, which could be termed the classically forbidden and
classically allowed regions, respectively. In the classically
forbidden region the behavior of the wave function is exponential,
while in the classically allowed region the wave function behaves
oscillatorily. In the quantum tunneling approach \cite{Vil}, the
wave function is so constructed as to create a universe emerging
from nothing by a tunneling procedure through a potential barrier in
the sense of usual quantum mechanics. Now, in our model, the
superpotential is always negative, which means that there is no
possibility of tunneling anymore, since a zero energy system is
always above the superpotential. In such a case, tunneling is no
longer required, as classical evolution is possible. As a
consequence, the wave function always exhibits oscillatory behavior.
In figure \ref{fig4}, we have plotted the square of the wave
functions for typical values of the parameters \footnote{In order to
make the physical predictions from a given wave function, one needs
to construct a probability measure. In quantum cosmology, since the
Wheeler-DeWitt equation is a Klein-Gordon type equation, a natural
choice is to deal with the conserved current ${\cal
J}=\frac{i}{2}(\Psi^*\nabla \Psi-\Psi \nabla \Psi^*)$, which
satisfies $\nabla.{\cal J}=0$. However, like the Klein-Gordon case
the probability measure constructed from this current is not
positive definite and its interpretation as a probability does not
work in a suitable manner. Because of such difficulties some authors
just consider the square of the wave function $|\Psi|^2$ as the
probability measure in the sense that the integral $\int_{\Omega}
|\Psi|^2 dV$ gives the probability of the universe being in the
region $\Omega$ of (mini)superspace. Although this definition has
also its own problems, it is excessively used in the minisuperspace
approximation of quantum cosmology \cite{Wil}.}. It is seen from
this figure that the wave function has a well-defined behavior near
$a=0$ and describes a universe emerging out of nothing without any
tunneling. Now to see that how the quantum solutions may describe an
expanding or contracting universe, we use a mechanism which we have
called the probabilistic evolutionary process (PEP), based on the
probabilistic structure of quantum systems, to provide a sense of
the evolution embedded in the wave function of the universe (see
\cite{Nima} for details). This is based on the fact that in quantum
systems the square of a state defines the probability, ${\cal
P}_a=|\Psi(a)|^2$. To make the discussion more clear, let us take a
specific initial condition corresponding to the point $P$. Then, PEP
states that the system (here specified by the scale factor $a$)
moves continuously to a state with higher probability and thus $P$
moves to the right to reach the point $Q$, a local maximum.
Therefore, the (expanding) universe begins its evolution by a
monotonically increasing scale factor to reach the point $Q$. In
this sense the transition $R\rightarrow Q$ describes a contracting
universe. On the other hand, the emergence of several peaks in the
wave function may be interpreted as a representation of different
quantum states that may communicate with each other through
tunneling. This means that there are different possible universes
(states) from which the present universe could have evolved and
tunneled in the past, from one universe (state) to another. Based on
this interpretation, one can argue that from the states located on
the larger peaks of the wave function a tunneling may occurs to
reach the states located on the smaller peaks and then the universe
evolved according to PEP. We schematically show such procedures as
\begin{eqnarray}\label{Z}
\mbox{expanding universe}: P{\buildrel {PEP} \over \longrightarrow
}Q{\buildrel {tunnel.} \over \longrightarrow} T{\buildrel {PEP}
\over \longrightarrow }U ...
\end{eqnarray}As the scale factor grows the probabilities become
small and smaller and the universe undergoes its classical region.
For the contraction universe similar discussion as above would be
applicable as well.

Now, let us to deal with the Wheeler-DeWitt equation (\ref{V}) in
the case of $k\neq 0$. Like the classical solutions in this case,
we analyze the quantum solutions in the limiting cases of small
and large scale factor. For the very small values for the scale
factor the terms with coefficients $g_c$ and $g_{\Lambda}$ are
negligible in comparison with the terms with HL parameters $g_r$
and $g_s$. In this limit the solutions to the equation (\ref{V})
read as
\begin{equation}\label{AB}
\Psi(a)=a\left[c_1J_{\sqrt{1-4kg_s}}\left(2k\sqrt{g_r}a\right)+c_2Y_{\sqrt{1-4kg_s}}\left(2k\sqrt{g_r}a\right)\right].
\end{equation}Since the behavior of the Bessel functions for the
small argument is $J_{\nu}(z)\sim z^{\nu}+O(z^{\nu+2})$ and
$Y_{\nu}(z)\sim z^{-\nu}+O(\frac{1}{z^{\nu(\nu-2)}})$, we set
$c_2=0$ and simplify the above relation for small $a$ as
\begin{equation}\label{AC}
\Psi(a)\sim a^{1+\sqrt{1-4kg_s}}.
\end{equation}The probability of creation an universe with scale
factor $a$ is
\begin{eqnarray}\label{AD}
|\Psi(a)|^2\sim \left\{
\begin{array}{ll}
a^2,\hspace{0.5cm}1-4kg_s<0,\\\\
a^{2+2\sqrt{1-4kg_s}},\hspace{0.5cm}1-4kg_s>0,
\end{array}\right.
\end{eqnarray}which in both cases describes an expanding non-singular
behavior in the early times of cosmic evolution in agreement with
the classical solutions presented in figure \ref{fig2}.

For the large scale factor, a WKB analysis like we have done for
the flat case yields the classical solutions (\ref{S})-(\ref{U}),
which shows that our treatment for quantization of the model lies
in a right way. However, for $a>>1$ we can neglect the last two
terms with coefficients $g_r$ and $g_s$ in (\ref{V}) and write its
solutions in terms of the Airy functions as
\begin{equation}\label{AE}
\Psi(a)=c_1
\mbox{Ai}\left(\frac{kg_c-g_{\Lambda}a^2}{g_{\Lambda}^{2/3}}\right)+c_2
\mbox{Bi}\left(\frac{kg_c-g_{\Lambda}a^2}{g_{\Lambda}^{2/3}}\right).
\end{equation}
\begin{figure}
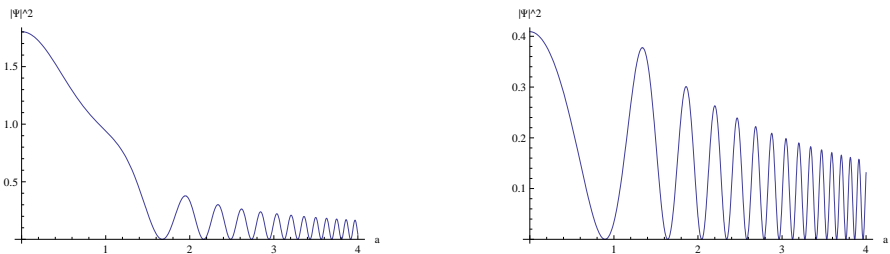

\begin{tabular}{c}\epsfig{figure=Fig7.eps,width=5cm}
\hspace{1.5cm} \epsfig{figure=Fig8.eps,width=5cm}
\end{tabular}
\caption{\footnotesize Left: Late square of the wave function
(\ref{AE}) for $k=1$. Right: The same figure for $k=-1$.
}\label{fig5}
\end{figure}A glance at the shape of this wave function which is
qualitatively plotted in figure \ref{fig5} for typical values of
the parameters, shows that it has a damping oscillatory behavior
denoting a classically allowed region. Such a wave function is
also found in \cite{Haw} to describe the dynamical behavior of a
universe dominated by the cosmological constant.
\section{Cosmological dynamics with perfect fluid}
\subsection{Classical model}
\begin{figure}
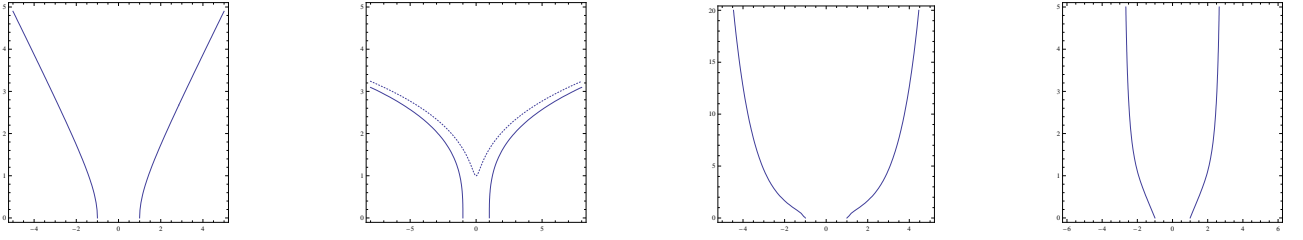

\begin{tabular}{c}\epsfig{figure=Fig9.eps,width=3cm}
\hspace{1.5cm} \epsfig{figure=Fig10.eps,width=3cm}\hspace{1.5cm}
\epsfig{figure=Fig11.eps,width=3cm}\hspace{1.5cm}
\epsfig{figure=Fig12.eps,width=3cm}
\end{tabular}
\caption{\footnotesize Qualitative behavior of the scale factor
versus time based on relation (\ref{AI}) with numerical values
$P_0=1$, $g_{\Lambda}=1$ and $t_0=1$ for the parameters. The
figures are plotted for $\omega=-1/3,-1,0,1/3$ from left to right.
The dashed line in the figure corresponding to the case
$\omega=-1$ shows the expectation value of the scale factor, see
(\ref{QL}).}\label{fig6}
\end{figure}

Now, we assume that a perfect fluid in its Schutz's representation
is coupled with gravity. In this case the Hamiltonian (\ref{M})
describes the dynamics of the system. The advantage of using
Schutz formalism is that, in a natural way, it can offer a time
parameter in terms of dynamical variables of the perfect fluid.
Indeed, the equations of motion for $T$ and $P_T$ read as
\begin{equation}\label{AF}
\dot{T}=\left\{T,H\right\}=\frac{N}{a^{3\omega}},\hspace{0.5cm}\dot{P_T}=\left\{P_T,H\right\}=0.
\end{equation}A glance at the above equations shows that with choosing
the gauge $N=a^{3\omega}$, we shall have
\begin{equation}\label{AG}
N=a^{3\omega}\Rightarrow T=t,
\end{equation}which means that variable $T$ may play the role of time in
the model. Therefore, the Friedmann equation $H=0$ can be written
in the gauge $N=a^{3\omega}$ as follows
\begin{equation}\label{AH}
\dot{a}^2=-g_cka^{6\omega}+g_{\Lambda}a^{6\omega+2}+g_rk^2a^{6\omega-2}+g_ska^{6\omega-4}+P_0a^{3\omega-1},
\end{equation}where we take $P_T=P_0=\mbox{const}.$ from the second equation
of equation (\ref{AF}). Like the previous section, let us deal
first with the solutions of this equation for the flat background,
$k=0$. In this case, the solutions can be represented by
\begin{equation}\label{AI}
\frac{2a^{\frac{3-3\omega}{2}}}{3\sqrt{P_0}(1-\omega)}\hspace{.4cm}F_{1\hspace{-.6cm}2}
\hspace{.4cm}\left(\frac{1-\omega}{2+2\omega},\frac{1}{2},\frac{3+\omega}{2+2\omega}
;-\frac{g_{\Lambda}}{P_0}a^{3+3\omega}\right)=\pm t-t_0,
\end{equation}for $\omega \neq 1$ (for $\omega=1$, there are no real solutions), where $F_{1\hspace{-.5cm}2}\hspace{.4cm}(a,b,c;z)$
is hypergeometric function and $t_0$ is a constant of integration.
In figure \ref{fig6} we have plotted the scale factor versus time
for several values of $\omega$. As is clear from the figures the
resulting cosmology is consisted of two contraction and expansion
branches which are separated from each other by a classically
forbidden region in which there are no physically acceptable
solutions.

For $k\neq 0$, it is better to deal with the equation (\ref{AH})
in some special cases. To this end, we shall choose typical values
for the equation of state parameter $\omega$ as follows.

$\bullet$ $\omega=-1/3$, for which equation (\ref{AH}) reads
\begin{equation}\label{AJ}
\dot{a}^2=-g_cka^{-2}+g_{\Lambda}+g_rk^2a^{-4}+g_ska^{-6}+P_0a^{-2}.
\end{equation}In the region where the scale factor is small, i.e.,
in the early times, we can keep only the terms with coefficients
$g_r$ and $g_s$ on the right hand side of the above equation which
results the following solution
\begin{equation}\label{AL}
\frac{(kg_ra^2-2g_s)\sqrt{g_ra^2+kg_s}}{2kg_r^2}=\pm t-t_0.
\end{equation}For large scale factors, on the other hand, the
other terms become more important and one obtains the late time
behavior as
\begin{equation}\label{AM}
a(t)=\left[g_{\Lambda}^2(\pm t-t_0)^2-(P_0-g_c k)\right]^{1/2}.
\end{equation}

$\bullet$ $\omega=1/3$, for which equation (\ref{AH}) takes the
form
\begin{equation}\label{AN}
\dot{a}^2=-g_cka^2+g_{\Lambda}a^4+g_ska^{-2}+P_0+g_rk^2,
\end{equation}which has the solutions
\begin{equation}\label{AO}
a(t)=\left(\pm \sqrt{g_s k}t-t_0\right)^{1/2},
\end{equation}for the early times, and
\begin{equation}\label{AP}
a(t)=\left(\pm t-t_0\right)^{-1},
\end{equation}for late times of cosmic evolution. A quick look at these
solutions shows that all of them have some kinds of singularities.
In the next subsection we will deal with the quantization of this
model to see how things change according to the quantum picture of
the corresponding cosmology.
\subsection{Quantum model}
Now, let us to investigate how the above picture may be modified
if one deals with the quantization of the model described by the
Hamiltonian (\ref{M}). The Wheeler-DeWitt equation corresponding
to this Hamiltonian reads
\begin{equation}\label{QA}
\left[a^{-1}\frac{\partial^2}{\partial
a^2}-a^{-2}\frac{\partial}{\partial
a}-4ia^{-3\omega}\frac{\partial}{\partial
T}+4\left(-g_cka+g_{\Lambda}a^3+\frac{g_rk^2}{a}+\frac{g_sk}{a^3}\right)\right]\Psi(a,T)=0,
\end{equation}in which we have taken the factor ordering parameter
$p=1$ as before. Separation the variables in the above equation in
the form
\begin{equation}\label{QB}
\Psi(a,T)=e^{iET}\psi(a),
\end{equation}yields
\begin{equation}\label{QC}
\left[a^2\frac{d^2}{da^2}-a\frac{d}{da}+4\left(-g_cka^4+g_{\Lambda}a^6+g_rk^2a^2+g_sk+Ea^{3-3\omega}\right)\right]\psi(a)=0.
\end{equation}If $k=0$, the above equation has exact solutions for
some special values of $\omega$ as
\begin{eqnarray}\label{QD}
\psi_E(a)=\left\{
\begin{array}{ll}
c_1\mbox{Ai}\left(-\frac{E+g_{\Lambda}a^2}{g_{\Lambda}^{2/3}}\right)+c_2\mbox{Bi}\left(-\frac{E+g_{\Lambda}a^2}{g_{\Lambda}^{2/3}}\right)
,\hspace{0.5cm}\omega=-1/3,\\\\
a\left[c_1J_{1/3}\left(\frac{2}{3}\sqrt{E+g_{\Lambda}}a^3\right)+c_2J_{-1/3}\left(\frac{2}{3}\sqrt{E+g_{\Lambda}}a^3\right)\right],
\hspace{0.5cm}\omega=-1,\\\\a^2e^{-\frac{2}{3}i\sqrt{g_{\Lambda}}a^3}\left[c_1M\left(\frac{5\sqrt{g_{\Lambda}}+2iE}{6\sqrt{g_{\Lambda}}},\frac{5}{3};
\frac{4}{3}i\sqrt{g_{\Lambda}}a^3\right)+c_2U\left(\frac{5\sqrt{g_{\Lambda}}+2iE}{6\sqrt{g_{\Lambda}}},\frac{5}{3};
\frac{4}{3}i\sqrt{g_{\Lambda}}a^3\right)\right],\hspace{0.5cm}\omega=0,
\end{array}\right.
\end{eqnarray}
where $M(a,b;z)$ and $U(a,b;z)$ are confluent hypergeometric
functions. Now the eigenfunctions of the Wheeler-DeWitt equation
can be written as
\begin{equation}\label{QE}
\Psi_E(a,T)=e^{iET}\psi_E(a).
\end{equation}We may now write the general solution to the Wheeler-DeWitt equation as a superposition of its eigenfunctions; that is,
\begin{equation}\label{QF}
\Psi(a,T)=\int_0^\infty A(E)\Psi_E(a,T)dE,
\end{equation}where $A(E)$ is a suitable weight function to construct the
wave packets. Since the above relations seem to be too complicated
to extract an analytical expression for the wave function, let us
focus our attention on the case $\omega=-1$ for which analytical
expression for the integral (\ref{QF}) is found if we choose the
function $A(E)$ to be a quasi-Gaussian weight factor. To end this,
we take $c_2=0$ to satisfy the boundary condition $\psi_E(a=0)=0$
and write down the wave function as
\begin{equation}\label{QG}
\Psi(a,T)=\int_0^\infty aA({\cal E})e^{-ig_{\Lambda}T}e^{i{\cal
E}T}J_{1/3}\left(\frac{2}{3}\sqrt{{\cal E}}a^3\right)d{\cal E},
\end{equation}where ${\cal E}=E+g_{\Lambda}$. Now, by using the
equality
\begin{equation}\label{QH}
\int_0^\infty e^{-\alpha z^2}z^{\nu+1}J_{\nu}(\beta
z)dz=\frac{\beta^{\nu}}{(2\alpha)^{\nu+1}}e^{-\frac{\beta^2}{4\alpha}},
\end{equation}we choose the weight function as $A({\cal E})={\cal E}^{1/6}e^{-\gamma {\cal
E}}$, to obtain
\begin{equation}\label{QI}
\Psi(a,T)={\cal
N}e^{-ig_{\Lambda}T}\frac{a^2}{(\gamma-iT)^{4/3}}\exp
\left(-\frac{1}{9}\frac{a^6}{\gamma-iT}\right),
\end{equation}where ${\cal N}$ is a numerical factor and $\gamma$
is an arbitrary positive constant. Now, having the above
expression for the wave function of the universe, we are going to
obtain the predictions for the behavior of the corresponding
cosmological dynamics. In general, one of the most important
features in quantum cosmology is the recovery of classical
cosmology from the corresponding quantum model or, in other words,
how can the Wheeler-DeWitt wave functions predict a classical
universe. In this approach, one usually constructs a coherent wave
packet with good asymptotic behavior in the minisuperspace,
peaking in the vicinity of the classical trajectory. On the other
hand, in an another approach to show the correlations between
classical and quantum pattern, following the many-worlds
interpretation of quantum mechanics \cite{tip}, one may calculate
the time dependence of the expectation value of a dynamical
variable $q$ as
\begin{equation}\label{QJ}
<q>=\frac{<\Psi|q|\Psi>}{<\Psi|\Psi>}.
\end{equation}Following this approach, we may write the expectation value for
the scale factor as
\begin{equation}\label{QK}
<a>(T)=\frac{\int_0^\infty
\Psi^{*}(a,T)a\Psi(a,T)da}{\int_0^\infty
\Psi^{*}(a,T)\Psi(a,T)da},
\end{equation}which upon substitution (\ref{QI}) yields
\begin{equation}\label{QL}
<a>(T)=a_0\left[\gamma^2+(g_{\Lambda}+P_0)T^2\right]^{1/6}.
\end{equation}This relation may be interpreted as the quantum counterpart
of the classical solutions (\ref{AI}) with $\omega=-1$. However,
in spite of the classical solutions, for the wave function
(\ref{QI}), the expectation value (\ref{QL}) of $a$ never
vanishes, showing that these states are nonsingular. Indeed, in
equation (\ref{QL}) $T$ varies from $-\infty$ to $+\infty$, and
any $T_0$ is just a specific moment without any particular
physical meaning like big-bang singularity. Now, let us take a
look at the $\omega=-1$ case of the figure \ref{fig6} in which the
expectation value (\ref{QL}) is plotted with the dashed line. As
is clear from this figure, for a perfect fluid with $\omega=-1$,
the corresponding classical cosmology admits two separate
solutions, which are disconnected from each other by a classically
forbidden region. One of these solutions represents a contracting
universe ending in a singularity while another describes an
expanding universe which begins its evolution with a big-bang
singularity. On the other hand, the evolution of the scale factor
based on the quantum mechanical considerations shows a bouncing
behavior in which the universe bounces from a contraction epoch to
a reexpansion era. Indeed, the classically forbidden region is
where the quantum bounce has occurred. We see that in the late
time of cosmic evolution in which the quantum effects are
negligible, these two behaviors coincide with each other. This
means that the quantum structure which we have constructed has a
good correlation with its classical counterpart.

For a background geometry with $k\neq 0$, to analyze the quantum
behavior of the model we may neglect the terms with coefficients
$g_c$, $g_{\Lambda}$ and $g_r$ in (\ref{QC}) in the early times,
i.e., in the region where the quantum effects have their dominate
role. In a such a situation this equation takes the form
\begin{equation}\label{QM}
\left[\frac{d^2}{da^2}-\frac{1}{a}\frac{d}{da}+4\left(\frac{g_sk}{a^2}+Ea^{1-3\omega}\right)\right]\psi(a)=0,
\end{equation}where its solution for some special cases are as
follows
\begin{eqnarray}\label{QN}
\psi_E(a)=\left\{
\begin{array}{ll}
aJ_{\frac{1}{2}\sqrt{1-4kg_s}}\left(\sqrt{E}a^2\right),
\hspace{0.5cm}\omega=-1/3,\\\\
aJ_{\sqrt{1-4kg_s}}\left(2\sqrt{E}a\right),
\hspace{0.5cm}\omega=1/3,
\end{array}\right.
\end{eqnarray}in which we have again applied the boundary
condition $\psi(a=0)=0$ on the eigenfunctions. Following the same
steps which led us to the wave function (\ref{QI}), we obtain the
wave function as
\begin{eqnarray}\label{QO}
\Psi(a,T)=\left\{
\begin{array}{ll}
\frac{a^{1+\sqrt{1-4kg_s}}}{(\gamma-iT)^{1+\frac{\sqrt{1-4kg_s}}{2}}}\exp\left[-\frac{a^4}{4(\gamma-iT)}\right],
\hspace{0.5cm}\omega=-1/3,\\\\
\frac{a^{1+\sqrt{1-4kg_s}}}{(\gamma-iT)^{1+\sqrt{1-4kg_s}}}\exp\left[-\frac{a^2}{\gamma-iT}\right],
\hspace{0.5cm}\omega=1/3,
\end{array}\right.
\end{eqnarray}from which the expectation values are obtained as
\begin{eqnarray}\label{QP}
<a>(T)\sim\left\{
\begin{array}{ll}
(\gamma^2+T^2)^{1/4},
\hspace{0.5cm}\omega=-1/3,\\\\
(\gamma^2+T^2)^{1/2}, \hspace{0.5cm}\omega=1/3.
\end{array}\right.
\end{eqnarray}The discussions on the comparison between quantum cosmological solutions and their
corresponding form from the classical formalism, i.e., equations
(\ref{AJ})-(\ref{AP}) are the same as previous model, namely the
flat model. Similar discussion as above would be applicable to
this case as well.
\section{Summary}
In this paper we have applied the recently proposed Ho\v{r}ava
theory of gravity to a FRW cosmological model. After a very brief
review of HL theory of gravity, we have considered a FRW
cosmological setting in the framework of the projectable HL
gravity without detailed balance condition and presented its
Hamiltonian in terms of the minisuperspace variables, both for the
vacuum and perfect fluid cases. For the flat model without the
matter contribution, we showed that the classical field equations
admit contracting and expanding de Sitter-like solutions in which
the cosmological constant is modified by the HL parameter
$\lambda$. For the non-flat background in this case, though the
corresponding Friedmann equation did not have exact solutions, we
analyzed the behavior of its solutions in the limiting cases of
the early and late times of cosmic evolution and obtained
analytical expressions for the scale factor in these regions. We
saw that these solutions are consisted of two separate branches
each of which exhibit some kinds of classical singularities. We
then have repeated the calculations when the model is augmented
with a perfect fluid as the matter field. Again, we showed that
the classical solutions have either contracting or expanding
branches which are disconnected from each other by some
classically forbidden regions. Another part of the paper is
devoted to the quantization of the model described above. For an
empty universe, we have shown that by applying the WKB
approximation on the Wheeler-DeWitt equation, one can recover the
late time behavior of the classical solutions. For the early
universe, we obtained oscillatory quantum states free of classical
singularities by which two branches of classical solutions may
communicate with each other. In the presence of matter, we focused
our attention on the approximate analytical solutions to the
Wheeler-DeWitt equation in the domain of small scale factor, i.e.
in the region which the quantum cosmology is expected to be
dominant. Using Schutz's representation for the perfect fluid,
under a particular gauge choice, we led to the identification of a
time parameter which allowed us to study the time evolution of the
resulting wave function. Investigation of the expectation value of
the scale factor shows a bouncing behavior near the classical
singularity. In addition to singularity avoidance, the appearance
of bounce in the quantum model is also interesting in its nature
due to prediction of a minimal size for the corresponding
universe. It is well-known that the idea of existence of a minimal
length in nature is supported by almost all candidates of quantum
gravity.


\begin{thebibliography}{99}
\bibitem{Hor1}P. Ho\v{r}ava, {\it Phys. Rev.} D {\bf 79} (2009) 084008 (arXiv: 0901.3775
[hep-th])\\P. Ho\v{r}ava, {\it Phys. Rev. Lett.} {\bf 102} (2009)
161301 (arXiv: 0902.3657 [hep-th])

\bibitem{Gian}G. Calcagni, {\it J. High Energy Phys.} {\bf JHEP 09} (2009)
112 (arXiv: 0904.0829 [hep-th])\\M.-I. Park, {\it J. High Energy
Phys.} {\bf JHEP 09} (2009) 123 (arXiv: 0905.4480 [hep-th])\\M.-I.
Park, {\it Class. Quantum Grav.} {\bf 28} (2011) 015004 (arXiv:
0910.1917 [hep-th])\\M. Minamitsuji, {\it Phys. Lett.} B {\bf 684}
(2010) 194 (arXiv: 0905.3892 [astro-ph])\\A. Wang and Y. Wu, {\it J.
Cosmol. Astropart. Phys.} {\bf JCAP 07} (2009) 012 (arXiv: 0905.4117
[hep-th])\\J. Greenwald, A. Papazoglou and A. Wang, {\it Phys. Rev.}
D {\bf 81} (2010) 084046 (arXiv: 0912.0011 [hep-th])\\E. Kiritsis
and G. Kofinas, {\it Nucl. Phys.} B {\bf 821} (2009) 467 (arXiv:
0904.1334 [hep-th])\\T.P. Sotiriou, {\it J. Phys. Conf. Ser.} {\bf
283} (2011) 012034 (arXiv: 1010.3218 [hep-th])\\D. Vernieri and T.P.
Sotiriou, {\it Phys. Rev.} D {\bf 85} (2012) 064003 (arXiv:
1112.3385 [hep-th])\\S. Mukohyama, {\it Class. Quantum Grav.} {\bf
27} (2010) 223101 (arXiv: 1007.5199 [hep-th])\\E.N. Saridakis, {\it
Aspects of Ho\v{r}ava-Lifshitz cosmology} (arXiv: 1101.0300
[astro-ph])\\M. Jamil and E.N. Saridakis, {\it J. Cosmol. Astropart.
Phys.} {\bf JCAP 07} (2010) 028 (arXiv: 1003.5637 [hep-th])\\O.
Obreg\'{o}n and J.A. Preciado, {\it Phys. Rev.} D {\bf 86} (2012)
063502

\bibitem{Blas}D. Blas, O. Pujolas and S. Sibiryakov, {\it Phys. Rev. Lett.} {\bf 104} (2010) 181302 (arXiv: 0909.3525
[hep-th])\\D. Blas, O. Pujolas and S. Sibiryakov, {\it J. High
Energy Phys.} {\bf JHEP 04} (2011) 018 (arXiv: 1007.3503 [hep-th])

\bibitem{Sot} T.P. Sotiriou, M. Visser and S. Weinfurtner, {\it Phys. Rev.
Lett.} {\bf 102} (2009) 251601 (arXiv: 0904.4464 [hep-th])\\
 T.P. Sotiriou, M. Visser and S. Weinfurtner, {\it J. High Energy Phys.} {\bf JHEP 10} (2009) 033
(arXiv: 0905.2798 [hep-th])

\bibitem{Ber}O. Bertolami and C.A.D. Zarro, {\it Ho\v{r}ava-Lifshitz Quantum Cosmology} (arXiv: 1106.0126
[hep-th])

\bibitem{Paulo}J. Paulo, M. Pitelli and A. Saa, {\it Phys. Rev.} D {\bf 86} (2012) 063506 (arXiv: 1204.4924
[gr-qc])

\bibitem{Chris}T. Christodoulakis and N. Dimakis, {\it Classical and Quantum Bianchi Type III vacuum Ho\v{r}ava-Lifshitz Cosmology} (arXiv: 1112.0903
[gr-qc])

\bibitem{Kei}K. Maeda, Y. Misonoh and T. Kobayashi, {\it Phys. Rev.} D {\bf 82} (2010) 064024 (arXiv: 1006.2739
[hep-th])

\bibitem{Schutz}B.F. Schutz, {\it Phys. Rev.} D {\bf 2} (1970) 2762\\B.F. Schutz, {\it Phys. Rev.} D {\bf 4} (1971)
3559\\V.G. Lapchinskii and V.A. Rubakov, {\it Theor. Math. Phys.}
{\bf 33} (1977) 1076

\bibitem{Vak}A.B. Batista, J.C. Fabris, S.V.B. Goncalves and J. Tossa, {\it Phys. Lett.} A {\bf 283} (2001) 62
(arXiv: gr-qc/0011102)\\ F.G. Alvarenga, J.C. Fabris, N.A. Lemos
and G.A. Monerat, {\it Gen. Rel. Grav.} {\bf 34} (2002) 651
(arXiv: gr-qc/0106051)\\ A.B. Batista, J.C. Fabris, S.V.B.
Goncalves and J. Tossa, {\it Phys. Rev.} D {\bf 65} (2002) 063519
(arXiv: gr-qc/0108053)\\B. Vakili, {\it Phys. Lett.} B {\bf 688}
(2010) 129 (arXiv: 1004.0306 [gr-qc])\\B. Vakili, {\it Class.
Quantum Grav.} {\bf 27} (2010) 025008 (arXiv: 0908.0998 [gr-qc])

\bibitem{Vil}A. Vilenkin, {\it Phys. Rev.} D {\bf 33} (1986) 3560\\A. Vilenkin, {\it Phys. Rev.} D {\bf 37} (1988) 888

\bibitem{Ch}T. Christodoulakis and J. Zanelli, {\it Nuovo Cim.} B {\bf 93}
(1986) 1

\bibitem{Wil} D. Wiltshire, {\it An Introduction to Quantum
Cosmology}, (arXiv: gr-qc/0101003)\\J.J. Halliwell, {\it
Introductory Lectures on Quantum Cosmology} (arXiv: 0909.2566
[gr-qc])

\bibitem{Nima} N. Khosravi, H.R. Sepangi and B. Vakili, {\it Gen.
Rel. Grav.} {\bf 42} (2010) 1081 (arXiv: 0909.2487 [gr-qc])

\bibitem{Haw}J.B. Hartle and S.W. Hawking, {\it Phys. Rev.} D {\bf 28} (1983) 2960

\bibitem{tip}F.J. Tipler, {\it Phys. Rep.} {\bf 137} (1986) 231

\end{thebibliography}
\end{document}